\documentclass[letterpaper]{article} 
\usepackage{aaai2026}  
\usepackage{times}  
\usepackage{helvet}  
\usepackage{courier}  
\usepackage[hyphens]{url}  
\usepackage{graphicx} 
\usepackage{natbib}  
\usepackage{caption} 
\usepackage{algorithm}
\usepackage{algorithmic}

\usepackage{newfloat}
\usepackage{listings}
\DeclareCaptionStyle{ruled}{labelfont=normalfont,labelsep=colon,strut=off} 
\floatstyle{ruled}
\newfloat{listing}{tb}{lst}{}
\floatname{listing}{Listing}
\usepackage{amssymb}
\usepackage{booktabs}       
\usepackage{amsmath}
\usepackage{tabularx}
\usepackage{multirow}
\usepackage{makecell}
\usepackage{hyperref}
\usepackage{xcolor}
\hypersetup{
    colorlinks=true,
    urlcolor=blue,
    citecolor=black,
    linkcolor=black,
}

\nocopyright

\title{HunyuanVideo-Foley: Multimodal Diffusion with Representation Alignment for High-Fidelity Foley Audio Generation}
\author{
    \normalsize
    Sizhe Shan\textsuperscript{\rm 1,2}\equalcontrib,
    Qiulin Li\textsuperscript{\rm 1,3}\equalcontrib,
    Yutao Cui\textsuperscript{\rm 1},
    Miles Yang\textsuperscript{\rm 1},
    Yuehai Wang\textsuperscript{\rm 2},
    Qun Yang\textsuperscript{\rm 3},
    Jin Zhou\textsuperscript{\rm 1}\thanks{Corresponding author},
    Zhao Zhong\textsuperscript{\rm 1}
}
\affiliations{
    \textsuperscript{\rm 1}Tencent Hunyuan,
    \textsuperscript{\rm 2}Zhejiang University,
    \textsuperscript{\rm 3}Nanjing University of Aeronautics and Astronautics\\

}

\begin{document}

\maketitle

\begin{abstract}
Recent advances in video generation produce visually realistic content, yet the absence of synchronized audio severely compromises immersion. To address key challenges in video-to-audio generation, including multimodal data scarcity, modality imbalance and limited audio quality in existing methods, we propose HunyuanVideo-Foley, an end-to-end text-video-to-audio framework that synthesizes high-fidelity audio precisely aligned with visual dynamics and semantic context. Our approach incorporates three core innovations: (1) a scalable data pipeline curating 100k-hour multimodal datasets through automated annotation; (2) a representation alignment strategy using self-supervised audio features to guide latent diffusion training, efficiently improving audio quality and generation stability; (3) a novel multimodal diffusion transformer resolving modal competition, containing dual-stream audio-video fusion through joint attention, and textual semantic injection via cross-attention. Comprehensive evaluations demonstrate that HunyuanVideo-Foley achieves new state-of-the-art performance across audio fidelity, visual-semantic alignment, temporal alignment and distribution matching. The demo page is available at: \url{https://szczesnys.github.io/hunyuanvideo-foley/}.

\end{abstract}

\section{Introduction}

Recent advances in video generation models \cite{polyak2025moviegencastmedia,gao2025seedance10exploringboundaries, kong2025hunyuanvideosystematicframeworklarge} have achieved notable success in synthesizing high-quality, photorealistic dynamic sequences. However, the absence of synchronized audio in these generated videos significantly undermines immersion. Traditional Foley art requires meticulous frame-by-frame creation by professionals, incurring substantial time and financial costs that render it incompatible with the efficiency of modern video generation systems. To address this limitation, researches on automated Foley generation have gained momentum.

Text-to-audio (TTA) synthesis constitutes an early approach to Foley generation, producing high-quality audio conditioned exclusively on textual descriptions. The state-of-the-art (SOTA) TTA methods can produce high-fidelity audio well-aligned with semantic descriptions. Nevertheless, restricted to textual guidance only, TTA methods cannot inherently generate audio aligned with video content, which is a critical requirement for Foley generation.

Video-to-Audio (V2A) generation aims to produce high-quality audio precisely synchronized with video, both semantically and temporally. Recent V2A approaches \cite{cheng2025mmaudiotamingmultimodaljoint,liu2025thinksoundchainofthoughtreasoningmultimodal} based on the Multimodal Diffusion Transformer (MMDiT) framework have shown significant progress. These methods leverage dual-modal inputs (video and text), utilizing pre-trained encoders to extract video features and text embeddings to guide audio synthesis through diffusion or flow-matching processes. However, existing V2A methods suffer from several key limitations. (1) {\bfseries Multimodal Data Scarcity}: Public datasets like VGGSound \cite{chen2020vggsoundlargescaleaudiovisualdataset} offer only $\sim$550 hours of low-quality video-audio pairs, while high-quality TTA datasets (e.g., AudioCaps \cite{audiocaps}, WavCaps \cite{Mei_2024}) lack video modality. The scarcity of multimodal data fundamentally limits the generalization capabilities of existing Text-Video-to-Audio (TV2A) models. (2) {\bfseries Modality Imbalance}: Current methods exhibit over-reliance on a single modality (typically text), often maintaining only coarse temporal alignment with video while inadequately responding to visual semantics. For instance, when processing text ``the sound of ocean waves" alongside video depicting a beach scene with people, seagulls and waves, the model exclusively generates the wave sounds while neglecting other audio elements (footstep sounds and seagull calls). This phenomenon demonstrates an imbalanced multimodal integration where textual cues dominate the audio generation at the expense of visual information. (3) {\bfseries Audio Quality}: The fidelity of audio generated by existing methods fails to meet professional standards, often exhibiting background noise and semantically inconsistent artifacts.

To overcome these challenges, we propose HunyuanVideo-Foley, an end-to-end multimodal TV2A generation model capable of synthesizing high-quality audio tightly aligned with both input video and text semantics. Our model adopts a multimodal flow-matching transformer paradigm trained on a large-scale text-video-audio multimodal dataset. First, to enable scalable multimodal dataset creation, we introduce a comprehensive data pipeline for automated labeling and filtering of collected data. This pipeline facilitated the construction of a $\sim$100k hours TV2A dataset. Second, to address modality imbalance, we propose a novel multimodal audio generation architecture comprising dual-stream MMDiT blocks and single-stream audio DiT blocks. The MMDiT incorporates joint self-attention with interleaved RoPE to strengthen temporal dependencies between video and audio, followed by the injection of textual information through the cross-attention mechanism. Third, we introduce a Representation Alignment (REPA) loss to enhance audio quality by aligning the hidden embeddings from the single-stream audio DiT block with the audio features extracted by a pre-trained self-supervised model. \cite{li2023selfsupervisedaudioteacherstudenttransformer}. Furthermore, we employ an enhanced autoencoder based on DAC \cite{kumar2023highfidelityaudiocompressionimproved}, replacing its discrete tokens into continuous 128-dimensional representations to significantly improve audio reconstruction capabilities. Our key contributions are summarized as follows:
\begin{itemize}
\item We propose an efficient TV2A data pipeline capable of automatically curating large-scale, high-quality datasets containing 100k-hour level text-video-audio pairs.
\item We introduce a REPA loss leveraging pre-trained audio features to provide semantic and acoustic guidance for the audio modeling process, effectively enhancing audio generation quality and stability.
\item We introduce HunyuanVideo-Foley, a novel TV2A framework that generates high-quality, semantically and temporally aligned audio from video and text inputs. Our approach mitigates modality imbalance, significantly enhancing visual-semantic alignment while sustaining text-semantic alignment, achieving SOTA performance.
\end{itemize}

\section{Related Work}

\subsection{Text-to-Audio}
Early audio synthesis focused on text-conditional generation, leveraging advancements in generative models. DiffSound \cite{yang2023diffsounddiscretediffusionmodel} pioneered diffusion models for environmental sound synthesis. AudioGen \cite{kreuk2023audiogentextuallyguidedaudio} adopts auto-regressive transformer to predict discrete audio representation. Subsequent advances including AudioLDM \cite{liu2023audioldmtexttoaudiogenerationlatent}, Make-An-Audio \cite{huang2023makeanaudiotexttoaudiogenerationpromptenhanced}, and Stable Audio Open \cite{evans2024stableaudioopen} utilized latent diffusion with text embeddings from pre-trained text encoders to enhance semantic alignment. Recently, TangoFlux \cite{hung2025tangofluxsuperfastfaithful} introduced a hybrid DiT architecture following Flux \cite{BlackForestLabsFlux2024}, with preference optimization, enabling high-fidelity TTA generation at reduced latency.

\subsection{Video-to-Audio}
Video-to-Audio synthesis aims to generate audio semantically and temporally consistent with video content. Existing V2A approaches can be broadly categorized into two paradigms, injecting visual features into pre-trained TTA models and training V2A models from scratch. In the first category, T2AV \cite{mo2024texttoaudiogenerationsynchronizedvideos} introduces an Audio-Visual ControlNet to strengthen visual consistency in TTA models. FoleyCrafter \cite{zhang2024foleycrafterbringsilentvideos} utilizes semantic adapters and temporal controllers for alignment, injecting textual and visual embeddings into a UNet backbone through cross-attention to guide audio generation. For approaches trained from scratch, Diff-Foley \cite{luo2023difffoleysynchronizedvideotoaudiosynthesis} utilizes a contrastive audio-visual pre-training (CAVP) module to align features across modalities. Recent works have demonstrated remarkable advances in both audio quality and multimodal alignment. Frieren \cite{wang2025frierenefficientvideotoaudiogeneration} proposes an efficient V2A model based on rectified flow matching. MMAudio \cite{cheng2025mmaudiotamingmultimodaljoint} adopts a hybrid architecture combining MMDiT blocks with single-modality DiT blocks, incorporating synchronization features via Synchformer \cite{iashin2024synchformerefficientsynchronizationsparse}, which is validated for temporal alignment efficacy in V-AURA \cite{viertola2024temporallyalignedaudiovideo}. MMAudio achieves high-quality synthesis with enhanced alignment. Concurrent work ThinkSound \cite{liu2025thinksoundchainofthoughtreasoningmultimodal} proposes a Chain-of-Thought (CoT) framework enabling step-by-step interactive audio generation and editing. 

While previous approaches have made significant progress in V2A synthesis, several critical challenges remain unresolved. These include suboptimal audio quality that falls short of professional standards, imprecise temporal alignment, and insufficient semantic correspondence with visual context. In contrast to existing methods, our approach employs distinct attention mechanisms to address the different alignment relationships between video-audio and text-audio modalities. This framework significantly enhances both video-semantic alignment and the quality of synthesized audio. 

\subsection{Representation Alignment}
Representation Alignment (REPA), first introduced by \cite{yu2024representation}, accelerates convergence and enhances semantic fidelity in large-scale generative models through aligning internal features with representations extracted from a pre-trained visual encoder. The REPA framework has since been widely adopted across generative modeling tasks. VA-VAE \cite{yao2025reconstruction} integrates REPA into LightningDiT to improve variational autoencoder latent space learning. JanusFlow \cite{ma2025janusflow} employs REPA for multimodal framework refinement; UniTok \cite{ma2025unitok} applies REPA to develop unified visual tokenizers; and MergeVQ \cite{li2025mergevq} utilizes REPA for vector quantization based model optimization. Building upon these successes, we apply REPA to TV2A synthesis, where we align intermediate representations of DiT blocks with frame-level audio features extracted from a pre-trained self-supervised model \cite{li2023selfsupervisedaudioteacherstudenttransformer} to enhance semantic and acoustic modeling. Our experimental results demonstrate marked improvements in both audio fidelity and semantic relevance.

\begin{figure*}[htbp]
    \centering    
    \includegraphics[width=0.7\textwidth]{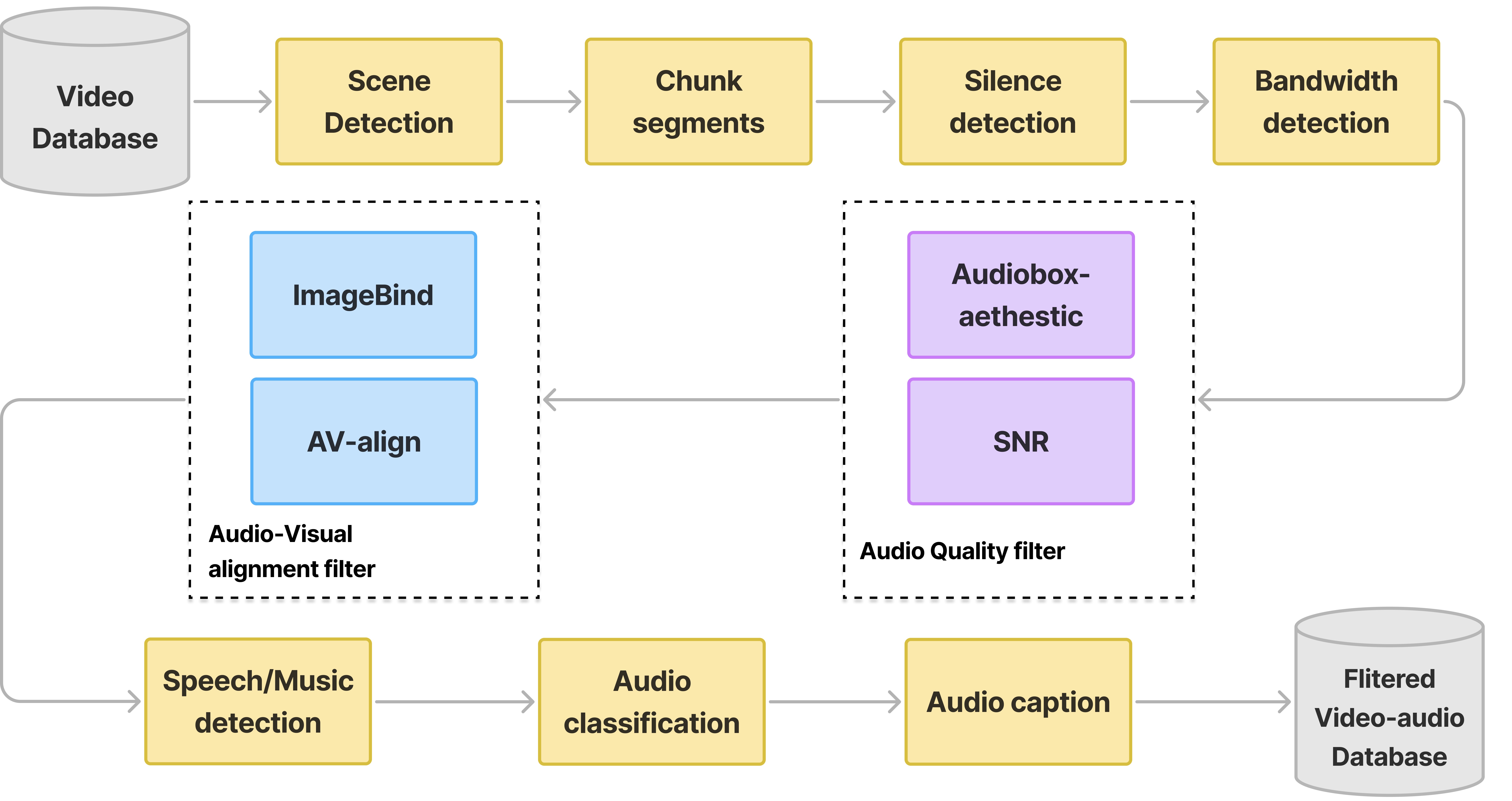}
    \caption{Data pipeline for filtering video-audio data. The workflow illustrates the processing steps from the raw video database to the filtered video-audio database.}
    \label{fig:data}
\end{figure*}
\section{Methodology}

\subsection{TV2A Data Pipeline}
The TV2A task presents a complex multimodal generation challenge that requires large-scale, high-quality text-video-audio datasets to produce robust and generalizable audio. Current open-source datasets, however, lack the necessary quality and scale to adequately support this demanding task. To bridge this gap, we develop a comprehensive data pipeline designed to systematically identify and exclude unsuitable content.

As illustrated in Figure \ref{fig:data}, our multi-stage filtering process firstly eliminate videos lacking audio streams. Subsequently, we employ scene detection algorithms to segment raw videos, then chunk them into 8-second intervals. These segments undergo silence ratio analysis, with those exceeding an 80\% silence threshold being discarded. Given the prevalence of heavily compressed and quality-degraded content on internet platforms, we implement bandwidth detection to ensure audio quality, retaining only samples with effective sampling rates exceeding 32 kHz. Audio quality constitutes a critical factor in generative audio tasks. Videos captured using substandard equipment often exhibit substantial background noise and ambient interference, rendering them unsuitable for generating cinematic-quality audio. To address this issue, we employed the AudioBox-aesthetic-toolkit \cite{tjandra2025metaaudioboxaestheticsunified} for audio quality assessment. Additionally, signal-to-noise ratio (SNR) measurements serve as supplementary metrics. Using these parameters, we empirically design a standard to filter and retain only high-quality audio segments. Another challenge in the V2A domain is ensuring audio-video alignment, which consists of both semantic and temporal alignment. We leverage ImageBind \cite{girdhar2023imagebindembeddingspacebind} and AV-align \cite{10.1609/aaai.v38i7.28486} to address the semantic and temporal alignments, respectively.

Following the aforementioned filtering process, we annotate the remaining video segments using speech-music detection and audio classification models. These annotations provide categorical tags for each segment, enabling effective management of category distribution and ensuring balanced representation in the training dataset. Subsequently, we generate audio captions for each segment using GenAU \cite{hajiali2025tamingdatatransformersaudio}, which provides concise descriptions of the audio content. Leveraging this data pipeline, we have constructed a high-quality TV2A dataset comprising $\sim$100k hours of text-video-audio material, providing robust support for model training.

\subsection{Overview of TV2A Framework}
\begin{figure*}[t]
    \centering    
    \includegraphics[width=0.75\textwidth]{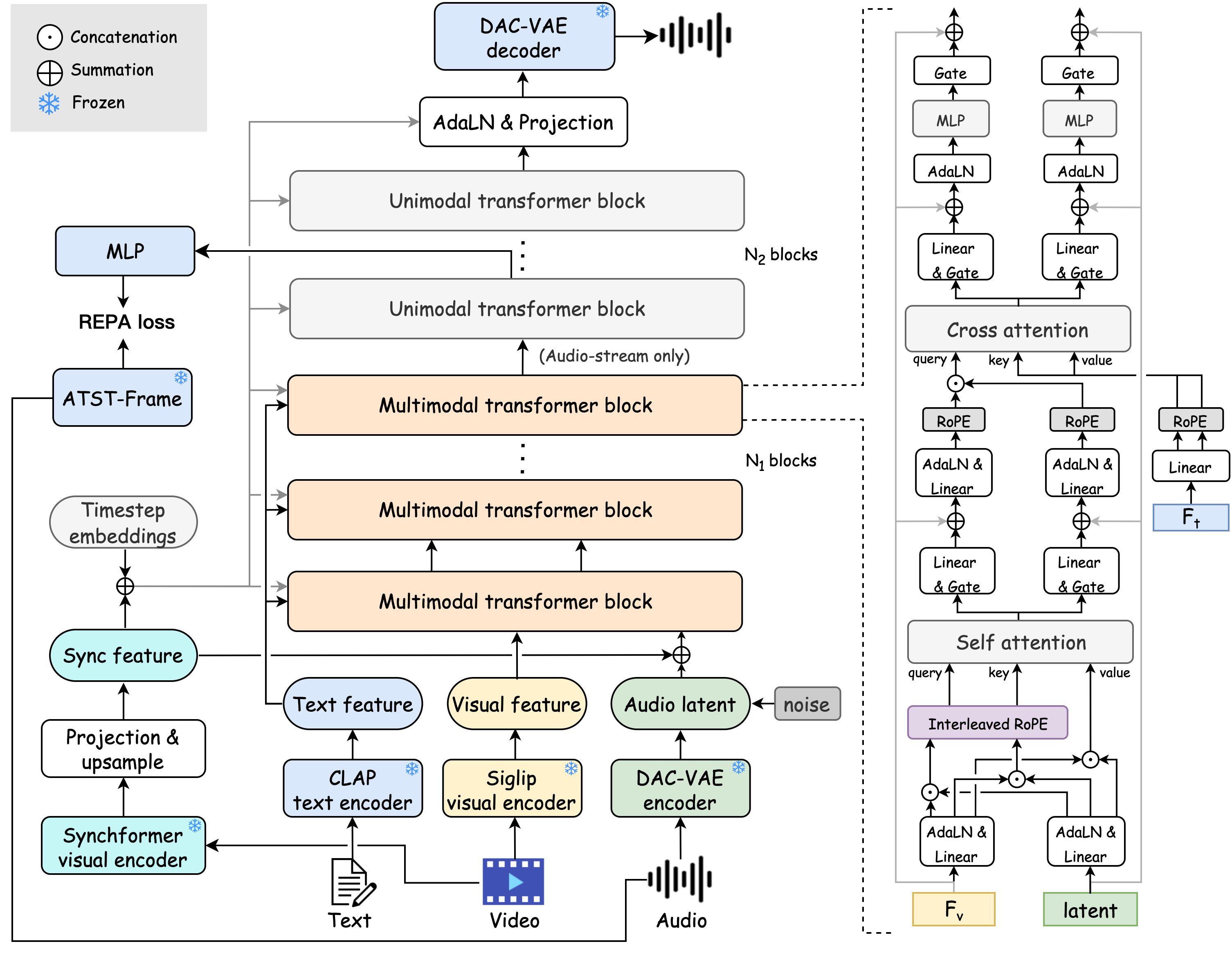}
    \caption{Overview of the HunyuanVideo-Foley model architecture. The proposed model integrates encoded text (CLAP), visual (SigLIP-2), and audio (DAC-VAE) inputs through a hybrid framework with $N_1$ multimodal transformer blocks followed by $N_2$ unimodal transformer blocks. The hybrid transformer blocks are modulated and gated with synchronization features and timestep embeddings. A pre-trained ATST-Frame is used to compute REPA loss with latnet representations from a unimodal transformer block. The generated audio latent are decoded into audio waveforms by the DAC-VAE decoder.}
    \label{fig:architecture}
\end{figure*}

To achieve modality balance and high-quality TV2A generation, we introduce the HunyuanVideo-Foley framework. As illustrated in Figure \ref{fig:architecture}, HunyuanVideo-Foley employs a hybrid architecture with $N_1$ multimodal transformer blocks (visual-audio streams) followed by $N_2$ unimodal transformer blocks (audio stream only). During training, video frames are encoded by a pre-trained visual encoder \cite{tschannen2025siglip2multilingualvisionlanguage} into visual features, while text captions are processed through a pre-trained text encoder \cite{10095889} to extract semantic features. Concurrently, raw audio undergoes an audio encoder to yield latent representations which are perturbed by additive Gaussian noise. The temporal alignment mechanism utilizes Synchformer-derived \cite{iashin2024synchformerefficientsynchronizationsparse} frame-level synchronization features to coordinate generation process through gated modulation pathways.

\subsection{Modality-Balanced MMDiT Architecture}

\subsubsection{Multimodal Representations.}
The multimodal input representations in HunyuanVideo-Foley leverage specialized encoders to extract modality-optimized features: video frames are processed by a pre-trained SigLIP2 \cite{tschannen2025siglip2multilingualvisionlanguage} encoder generating visual features $F_v \in \mathbb{R}^{B \times L_v\times D}$. Textual descriptions are encoded through CLAP \cite{10095889}, yielding semantic embeddings $F_t \in \mathbb{R}^{B \times L_t\times D}$. Audio waveforms are compressed via our enhanced DAC-VAE encoder into audio latents $x \in \mathbb{R}^{B \times L_a\times D}$. Synchronization features $F_s \in \mathbb{R}^{B \times L_s\times D}$ are extracted using Synchformer \cite{viertola2024temporallyalignedaudiovideo}, where $B$ denotes batch size, $L_v$, $L_t$, $L_a$, and $L_s$ represent the correspondence of sequence lengths, and $D$ is the feature dimension.

\subsubsection{Multimodal Alignment with Dual-Phase Attentions.}
TV2A generation requires modeling distinct alignment relationships between modalities: audio and video exhibit frame-level temporal dependencies, while audio-text interactions rely on global semantic guidance.
To address this dichotomy, our MMDiT architecture employs a dual-phase attention mechanism. In self-attention blocks, audio latents and visual features are concatenated into a unified sequence, enhanced with interleaved RoPE. These fused sequences serve as query, key, and value, capturing precise synchronization relations. The aligned audio-visual features then split into parallel streams, each processed through linear projection layers with gating mechanisms, followed by adaptive layer normalization (adaLN) layers that dynamically modulate both audio latents and visual features using the synchronization features from Synchformer. In the following cross-attention blocks, the concatenated audio-visual sequence acts as query, while CLAP-derived text embeddings provide key and value.

\subsubsection{Interleaved RoPE.}
To enhance the temporal alignment between audio and visual modalities in MM-DiT, we adopt an interleaved rotary position embedding (RoPE) mechanism. Traditional approaches apply separate RoPE \cite{SU2024127063} encodings to audio and visual sequences independently, which may not effectively capture the temporal correlations. Our interleaved RoPE strategy interleaves audio and visual tokens along the temporal dimension before applying position embeddings, thereby enabling the model to learn more coherent temporal relationships between visual-audio modalities. Specifically, given audio latents $x \in \mathbb{R}^{B \times L_a \times H \times D}$ and visual features $F_v \in \mathbb{R}^{B \times L_v \times H \times D}$, where $B$ denotes batch size, $L_a$ and $L_v$ represent sequence lengths, $H$ is the number of attention heads, and $D$ is the head dimension. We first employ nearest-neighbor interpolation to align the sequence length to $L=max(L_a,L_v)$, then interleave two sequences. For timestep $t \in [1,L]$, the joint feature $F_{av}$ is combined through Equation \eqref{eq:1}:
\begin{equation}
\label{eq:1}
\begin{cases}
\begin{aligned}
F_{av}[&\colon, 2t-1, \colon, \colon] = x[\colon, t, \colon, \colon] \\
F_{av}[&\colon, 2t, \colon, \colon] = F_v[\colon, t, \colon, \colon]
\end{aligned}
\end{cases}
\end{equation}
This operation creates an alternating pattern $F_{av}$ of audio and visual tokens. Subsequently, we apply RoPE to this interleaved sequence, ensuring that temporally adjacent audio and visual tokens receive consecutive position embeddings. Finally, we decouple the interleaved sequence back into separate audio and visual components, and then concatenate two sequences along the temporal dimension to serve as the query or key in the following self-attention mechanism. The interleaved RoPE ensures that the model can effectively capture the inherent temporal structure between audio latents and visual features, leading to improved generation quality and temporal coherence in the generation process. 

\subsubsection{Modulation with Synchronization Features.}
The model implements a dynamic conditioning scheme combining modulation and gated layers. The conditioning signal $\mathbf{c}$ is formulated as the summation of synchronization features and flow timestep embeddings, shown in Equation \eqref{eq:mod1}:
\begin{equation}
\label{eq:mod1}
\mathbf{c} = \mathbf{c}_{\text{sync}} + \mathbf{c}_{t}
\end{equation}
where $\mathbf{c}_{\text{sync}}$ encodes audio-visual temporal correlations and $\mathbf{c}_{t}$ represents the progression of the flow-matching process. To obtain modulation parameters, the composite signal $c$ is processed through parallel transformation pathways, as shown in Equation \eqref{eq:mod2}:

\begin{equation}
\resizebox{0.9\linewidth}{!}{
  $\begin{array}{lll}
  \alpha = W_\alpha \cdot \text{SiLU}(\mathbf{c}); &
  \beta = W_\beta \cdot \text{SiLU}(\mathbf{c}); &
  g = W_g \cdot \text{SiLU}(\mathbf{c})
  \end{array}$
}
\label{eq:mod2}
\end{equation}
where $\alpha$ and $\beta$ are two parameters of adaLN, $g$ is the parameter of gate mechanism, and weight matrices $W_\alpha, W_\beta, W_g$ is initialized to zero vectors to ensure stable convergence. Within each transformer block, input features $\mathbf{y}$ first undergo adaLN, shown in Equation \eqref{eq:mod3}:
\begin{equation}
\label{eq:mod3}
\mathbf{y}_{\text{mod}} = \text{LayerNorm}(\mathbf{y}) \odot (\mathbf{1} + \alpha) + \beta
\end{equation}
where $\odot$ denotes element-wise multiplication. The modulated features subsequently undergo attention computation, where the attention outputs are adaptively gated and integrated through residual connections, shown in Equation \eqref{eq:mod4}:
\begin{equation}
\label{eq:mod4}
\mathbf{y}_{\text{out}} = \mathbf{y}+\mathbf{y}_{\text{attn}} \odot g
\end{equation}
This conditioning mechanism ensures temporal coherence is maintained at both multimodal interaction and unimodal processing stages.

\subsection{REPA Training Strategy}
In the HunyuanVideo-Foley framework, we introduce the REPA Loss that involves aligning hidden states from intermediate layers of transformer blocks in our diffusion model with frame-level audio representations from the pre-trained ATST-Frame encoder.

Specifically, let $E_{ATST}$ denote the pre-trained encoder, which produces representations $\mathbf{F_r} \in \mathbb{R}^{B \times L_r\times D}$, where $\mathbf{F_r} = E_{ATST}(\mathbf{x})$, where $B$ is the batch size, $L_r$ representing the sequence length of audio feature and $D$ is the feature dimension. REPA loss aims to align the mapped latents from the intermediate DiT layers $\mathbf{h}$ with the ATST-Frame audio features $\mathbf{F_r}$. This mapping is implemented through a Multi-Layer Perceptron (MLP), such that $\mathbf{H} = \text{MLP}(\mathbf{h})$. Equation \eqref{eq:repa} shows the calculation of REPA loss.
\begin{equation}
\label{eq:repa}
\mathcal{L}_{REPA} = -\frac{\mathbf{F_r} \cdot \mathbf{H}}{|\mathbf{F_r}| \cdot |\mathbf{H}|}
\end{equation}
By maximizing the cosine similarity between the pre-trained representations and the internal representations from DiT layers, REPA Loss enables more effective semantic and acoustic guidance during audio generation modeling, thereby enhancing the semantic alignment and the quality of generated audio.

\begin{table*}[htb]
\small
\caption{Objective evaluation results on Kling-Audio-Eval. HunyuanVideo-Foley achieves superior performance across distribution matching (FD\textsubscript{PaNNs}, KL), audio quality (PQ), visual-semantic alignment (IB) and temporal alignment (DeSync) metrics.}
\centering
\label{tab:1}
\resizebox{0.85\textwidth}{!}{
\begin{tabular}{lllllllllllll}
\toprule
Method&FD\textsubscript{PaNNs}$\downarrow$&FD\textsubscript{PaSST} $\downarrow$&KL $\downarrow$ &IS $\uparrow$ &PQ $\uparrow$  &PC $\downarrow$&CE $\uparrow$&CU $\uparrow$ &IB $\uparrow$ & DeSync $\downarrow$ &CLAP $\uparrow$\\
\midrule
FoleyCrafter&22.30&322.63&2.47&7.08&6.05&2.91&3.28&5.44&0.22&1.23&0.22 \\
V-AURA&33.15&474.56&3.24&5.80&5.69&3.98&3.13&4.83&0.25&0.86&0.13  \\
Frieren&16.86&293.57&2.95&7.32&5.72&{\bfseries2.55}&2.88&5.10&0.21&0.86&0.16  \\
MMAudio (L-44.1kHz)&9.01&205.85&2.17&{\bfseries9.59}&5.94&2.91&{\bfseries3.30}&5.39&0.30&0.56&{\bfseries0.27} \\
ThinkSound ($w/o.$ CoT)&9.92&228.68&2.39&6.86&5.78&3.23&3.12&5.11&0.22&0.67&0.22  \\
HunyuanVideo-Foley (ours)&{\bfseries6.07}&{\bfseries202.12}&{\bfseries1.89}&8.30&{\bfseries6.12}&2.76&3.22&{\bfseries5.53}&{\bfseries0.38}&{\bfseries0.54}&0.24\\
\bottomrule
\end{tabular}
}
\end{table*}

\begin{table*}[htb]
\small
\caption{Objective evaluation results on VGGSound-Test. Our models achieves superior performance across audio quality (PQ) and visual-semantic alignment (IB).}
\centering
\label{tab:2}
\resizebox{0.85\textwidth}{!}{
\begin{tabular}{lllllllllllll}
\toprule
Method&FD\textsubscript{PaNNs}$\downarrow$&FD\textsubscript{PaSST} $\downarrow$&KL $\downarrow$ &IS $\uparrow$ &PQ $\uparrow$  &PC $\downarrow$&CE $\uparrow$&CU $\uparrow$ &IB $\uparrow$ & DeSync $\downarrow$ &CLAP $\uparrow$\\
\midrule
FoleyCrafter&20.65&171.43&2.26&14.58&6.33&2.87&3.60&5.74&0.26&1.22&0.19 
\\
V-AURA&18.91&291.72&2.40&8.58&5.70&4.19&3.49&4.87&0.27&0.72&0.12  \\
Frieren& 11.69&83.17&2.75&12.23&5.87&2.99&3.54&5.32&0.23&0.85&0.11 \\
MMAudio (L-44.1kHz)&{\bfseries7.42}&116.92&{\bfseries1.77}&{\bfseries21.00}&6.18&3.17&{\bfseries4.03}&5.61&0.33&{\bfseries0.47}&{\bfseries0.25} \\
ThinkSound ($w/o.$ CoT)&8.46&{\bfseries67.18}&1.90&11.11& 5.98&3.61&3.81&5.33&0.24&0.57&0.16 \\
HunyuanVideo-Foley (ours)& 11.34&145.22&2.14&16.14&{\bfseries6.40}&{\bfseries2.78}&3.99&{\bfseries5.79}&{\bfseries0.36}&0.53&0.24 \\
\bottomrule
\end{tabular}
}
\end{table*}

\begin{table*}[h!]
\small
\caption{Objective and subjective evaluation results on MovieGen-Audio-Bench. Our model achieves SOTA performance across almost all objective metrics and subjective evaluations.}
\centering
\label{tab:3}
\resizebox{0.85\textwidth}{!}{
\begin{tabular}{llllllllllll}
\toprule
Method& PQ $\uparrow$  &PC $\downarrow$&CE $\uparrow$&CU $\uparrow$ &IB $\uparrow$& DeSync $\downarrow$ &CLAP $\uparrow$ & MOS-Q$\uparrow$&MOS-S $\uparrow$&MOS-T $\uparrow$\\
\midrule
FoleyCrafter& 6.27 & {\bfseries2.72} & 3.34 & 5.68 & 0.17 & 1.29 & 0.14&3.36$\pm$0.78&3.54$\pm$0.88&3.46$\pm$0.95\\
V-AURA& 5.82 & 4.30 & 3.63 & 5.11 & 0.23 & 1.38 & 0.14&2.55$\pm$0.97&2.60$\pm$1.20&2.70$\pm$1.37 \\
Frieren& 5.71 & 2.81 & 3.47 & 5.31 & 0.18 & 1.39 & 0.16&2.92$\pm$0.95&2.76$\pm$1.20&2.94$\pm$1.26 \\
MMAudio (L-44.1kHz)& 6.17 & 2.84 & 3.59  & 5.62  & 0.27 & 0.80& {\bfseries0.35}&3.58$\pm$0.84&3.63$\pm$1.00&3.47$\pm$1.03\\
ThinkSound ($w/o.$ CoT)& 6.04 & 3.73 & 3.81 & 5.59 & 0.18 & 0.91 & 0.20&3.20$\pm$0.97&3.01$\pm$1.04&3.02$\pm$1.08 \\
HunyuanVideo-Foley (ours)&{\bfseries6.59} & 2.74 & {\bfseries3.88} & {\bfseries6.13} & {\bfseries0.35} & {\bfseries0.74}& 0.33&{\bfseries4.14$\pm$0.68}&{\bfseries4.12$\pm$0.77}&{\bfseries4.15$\pm$0.75}\\
\bottomrule
\end{tabular}
}
\end{table*}

\section{Experiments}

\subsection{Experiment Settings}

\subsubsection{Autoencoder.}
In the autoencoder framework, we develop DAC-VAE by replacing the Residual Vector Quantization (RVQ) blocks in DAC with a variational autoencoder architecture. This approach employs Gaussian distribution modeling in the latent space and substitutes quantization loss with KL divergence regularization, thereby achieving continuous encoding. Our DAC-VAE is trained on approximately 100k hours of audio data for 700k steps using 32 NVIDIA H20 GPUs with a batch size of 256. We adopt the AdamW optimizer with a learning rate of 1e-4 for optimization. The implemented system operates at a sampling rate of 48kHz, with a latent vector dimensionality of 128 and a latent rate of 50Hz.

\subsubsection{Implementation Details.}
HunyuanVideo-Foley consists of 18 MMDiT layers and 36 unimodal audio DiT layers with a hidden dimension of 1536 and 12 attention heads. The training is conducted on 128 H20 GPUs with an effective batch size of 2048 over 200k steps on a 100k-hour level TV2A datasets built by our proposed data pipeline, using the AdamW optimizer with a learning rate of 1e-4. We applied a classifier-free guidance (CFG) dropout rate of 0.1 for each modality. For evaluation, we conduct objective metrics comparisons between HunyuanVideo-Foley and existing SOTA models across Kling-Audio-Eval \cite{wang2025klingfoleymultimodaldiffusiontransformer}, VGGSound-Test set \cite{chen2020vggsoundlargescaleaudiovisualdataset}, and MovieGen-Audio-Bench \cite{polyak2025moviegencastmedia}. Additionally, we perform subjective tests on MovieGen-Audio-Bench to assess perceptual quality through human evaluations.

\subsubsection{Evaluation Metrics.}
For comprehensive evaluation, we adopt a multi-dimensional suite of metrics, assessing critical dimensions: {\bfseries distribution matching} via Fréchet Distance (FD) and Kullback-Leibler divergence (KL) using PANNs \cite{kong2020pannslargescalepretrainedaudio} and PaSST \cite{Koutini_2022} as feature extractors; {\bfseries audio quality} measured through the Inception Score (IS) computed with the PANNs classifier alongside AudioBox-Aesthetics comprising Production Quality (PQ), Production Complexity (PC), Content Enjoyment (CE) and Content Usefulness (CU); {\bfseries visual-semantic alignment} quantified by ImageBind (IB) measuring cosine similarity between input video and generated audio embeddings; {\bfseries temporal alignment} evaluated via DeSync predicted by Synchformer; {\bfseries text-semantic consistency} assessed by the LAION-CLAP score \cite{wu2024largescalecontrastivelanguageaudiopretraining}. For subjective evaluation, we adopt the mean opinion score (MOS) to evaluate audio quality (MOS-Q), semantic alignment (MOS-S), and temporal alignment (MOS-T). In audio reconstruction, we adopt Perceptual Evaluation of Speech Quality (PESQ), Short-Time Objective Intelligibility (STOI), Scale-Invariant Signal-to-Distortion Ratio (SI-SDR) and Mel distance.

\subsection{Main Results}

\subsubsection{Text-Video-to-Audio Generation.}

Table \ref{tab:1} presents the objective evaluation results on the Kling-Audio-Eval dataset. HunyuanVideo-Foley demonstrates superior performance across multiple metrics, including distribution matching (FD, KL), audio quality (PQ), visual-semantic alignment (IB), and temporal synchronization (DeSync) in comparison with baselines. Compared with the current state-of-the-art model MMAudio, HunyuanVideo-Foley demonstrates slightly inferior performance on IS, CE, and CLAP scores, while achieving notable improvements in FD (9.01 to 6.07), KL (2.17 to 1.89), and IB (0.30 to 0.38) scores.

The objective evaluation on the VGGSound-Test is shown in Table \ref{tab:2}. Notably, HunyuanVideo-Foley underperforms some baselines in distribution matching metrics (FD, KL), but leads in audio quality metrics (IS, PQ). This discrepancy may stem from the fact that most audio samples in VGGSound are recorded using non-professional equipment, resulting in generally poor audio quality that creates a substantial distribution gap with the outputs of HunyuanVideo-Foley. Nevertheless, our model maintains the SOTA performance in IB score while achieving comparable results in DeSync and CLAP metrics.

\begin{table*}[h!]
\small
\caption{Evaluation of autoencoder reconstructions in AudioSet (Sound), Song Describer (Music) and LibriTTS (Speech).}
\centering
\label{tab:4}
\resizebox{0.85\textwidth}{!}{
\begin{tabular}{lllllllllllll}
\toprule
&Dataset&Method&Sample rate&PESQ$\uparrow$&STOI $\uparrow$ &SI-SDR $\uparrow$&Mel-dist $\downarrow$ & Latent rate & Latent\\
\midrule
& \multirow{4}{*}{AudioSet}&DAC & 44.1kHz&4.17&0.94&11.08&0.48&86Hz&discrete \\ 
&&Stable Audio Open&44.1kHz&2.33&0.72&3.32&0.83&21.5Hz&64-dim \\
&&DAC-VAE (ours)&48kHz&3.59&0.91&8.41&0.60&50Hz&64-dim  \\
&&DAC-VAE (ours)&48kHz&{\bfseries4.45}&{\bfseries0.98}&{\bfseries14.76}&{\bfseries0.27}&50Hz&128-dim  \\
\midrule
&\multirow{4}{*}{\makecell[l]{Song\\Describer\\}}&DAC & 44.1kHz&4.18&0.96&13.84&0.48&86Hz&discrete \\ 
&&Stable Audio Open&44.1kHz &2.56&0.83&8.02&0.79&21.5Hz&64-dim \\
&&DAC-VAE (ours)&48kHz&3.57&0.93&12.60&0.57&50Hz&64-dim  \\
&&DAC-VAE (ours)&48kHz&{\bfseries4.45}&{\bfseries0.99}&{\bfseries17.40}&{\bfseries0.29}&50Hz&128-dim  \\ 
\midrule
&\multirow{4}{*}{\makecell[l]{LibriTTS\\Clean Set\\}}&DAC& 44.1kHz&4.29&0.98&12.37&0.47&86Hz&discrete \\ 
&&Stable Audio Open&44.1kHz &2.68&0.93&5.78&0.87&21.5Hz&64-dim \\
&&DAC-VAE (ours)&48kHz&3.75&0.97&9.51&0.61&50Hz&64-dim  \\
&&DAC-VAE (ours)&48kHz&{\bfseries4.50}&{\bfseries0.99}&{\bfseries14.37}&{\bfseries0.26}&50Hz&128-dim  \\ 
\bottomrule
\end{tabular}
}
\end{table*}

Table \ref{tab:3} displays both objective and subjective evaluation results on the MovieGen-Audio-Bench. HunyuanVideo-Foley exhibits outstanding generation quality, outperforming baselines in nearly all objective metrics and all subjective evaluations. Compared with the strong baseline MMAudio, our model demonstrates significant improvements across audio quality (PQ), temporal alignment (DeSync), and visual-semantic alignment (IB), while maintaining comparable performance in text-semantic alignment (CLAP).

Comprehensive evaluation across all three datasets demonstrates that HunyuanVideo-Foley achieves substantial improvements in visual-semantic alignment (IB) over all baselines. Our model also leads in audio quality (PQ) and temporal alignment (DeSync) while maintaining competitive text semantic alignment (CLAP). In terms of distribution matching, HunyuanVideo-Foley achieves optimal performance on the Kling-Audio-Eval dataset. These results collectively demonstrate that HunyuanVideo-Foley establishes new state-of-the-art performance in TV2A generation.

\subsubsection{Audio Reconstruction.}
For audio reconstruction, we conduct comparative studies between DAC \cite{kumar2023highfidelityaudiocompressionimproved} and the continuous VAE employed in Stable Audio Open \cite{evans2024stableaudioopen}. The evaluation spanned three distinct domains: AudioSet for general sounds, Song Describer for music, and LibriTTS-Clean testset for speech scenarios. As shown in Table \ref{tab:4}, our proposed DAC-VAE achieves superior performance across all metrics on three evaluation sets. These experiments validate that our DAC-VAE delivers robust reconstruction performance across diverse audio domains, establishing its effectiveness as a general-purpose audio reconstruction framework. 

\begin{table*}[h!]
\small
\caption{Ablation Study on Multimodal Transformer Block Architectures.}
\centering
\label{tab:5}
\resizebox{0.75\textwidth}{!}{
\begin{tabular}{llllllll}
\toprule
Method& PQ $\uparrow$  &PC $\downarrow$&CE $\uparrow$&CU $\uparrow$ &IB $\uparrow$ & DeSync $\downarrow$ &CLAP $\uparrow$ \\
\midrule
Joint self-attention &6.32&{\bfseries2.72}&3.61&5.76&0.31&1.05&{\bfseries0.32} \\
Parallel cross-attention &6.33&2.80&3.59&5.55&0.26&0.81&0.27  \\
\midrule
Joint self-attention+cross-attention &{\bfseries6.38}&2.76&{\bfseries3.68}&{\bfseries5.90}&{\bfseries0.32}&{\bfseries0.78}&0.30 \\
$w/o.$ interleaved RoPE& 6.36&2.78&3.65&5.77&0.31&0.79&0.30 \\
$w/o.$ unimodal DiT &6.23&2.83&3.57&5.70&0.31&0.79&0.30 \\
\bottomrule
\end{tabular}
}
\end{table*}

\begin{table}[h!]
\small
\caption{Ablation Study on Representation Alignment models.}
\centering
\label{tab:6}
\resizebox{0.48\textwidth}{!}{
\begin{tabular}{llllllll}
\toprule
Method& PQ $\uparrow$  &PC $\downarrow$&CE $\uparrow$&CU $\uparrow$ &IB $\uparrow$ & DeSync $\downarrow$ &CLAP $\uparrow$ \\
\midrule
$w/o.$ REPA & 6.23&2.83&3.57&5.63&0.31&0.79&0.30\\
EAT+ATST& 6.00& 2.90& 3.54 & 5.43& 0.32& 0.79& 0.29\\
EAT only & 6.24& 2.77& 3.55&{\bfseries5.69}&0.32&0.79&0.31 \\
ATST only & {\bfseries6.28}&{\bfseries2.74}&{\bfseries3.59}&5.68&{\bfseries0.33}&{\bfseries0.75}&{\bfseries0.34} \\
\bottomrule
\end{tabular}}
\end{table}

\begin{table}[h!]
\small
\caption{Ablation Study on Representation Alignment for Multimodal and Unimodal Transformers.}
\centering
\label{tab:7}
\resizebox{0.48\textwidth}{!}{
\begin{tabular}{llllllll}
\toprule
Method& PQ $\uparrow$  &PC $\downarrow$&CE $\uparrow$&CU $\uparrow$ &IB $\uparrow$ & DeSync $\downarrow$ &CLAP $\uparrow$ \\
\midrule
MMDiT + UniDiT &6.28&2.74&3.59&5.68&0.33&0.75&{\bfseries0.33}\\
MMDiT only &6.28&2.79&{\bfseries3.70}&5.73&0.33&0.79&0.32\\
\midrule
UniDiT only (Layer 8) &{\bfseries6.34}&2.80&3.67&{\bfseries5.77}&{\bfseries0.34}&{\bfseries0.74}&{\bfseries0.33}  \\
\midrule
Layer 12 &6.28&{\bfseries2.84}&3.61&{\bfseries5.77}&0.32&0.81&0.32  \\
Layer 16 &6.32&2.74&3.58&5.75&0.33&0.78&0.32 \\
\bottomrule
\end{tabular}}
\end{table}

\subsection{Ablation Study}
To thoroughly investigate the impact of different model architectures on performance and validate the effectiveness of the proposed design, we conduct meticulous ablation experiments on MovieGen-Audio-Bench. The ablation study primarily focus on multimodal conditioning methods in MMDiT, the efficacy of unimodal audio DiT, and optimal implementation strategies for representation alignment.
\subsubsection{Model Architecture.}

For the architecture of MMDiT, we design two alternative experiments: (1) employing joint self-attention for text-audio-video triple-stream modal alignment, and (2) using parallel cross-attention to separately align audio-text and audio-video modals. All configurations maintain identical experimental setups with excluding REPA and employing unimodal DiT. As shown in Table \ref{tab:5}, the proposed approach, which first achieves audio-video alignment through joint attention, and then injects text features through cross-attention to the audio-video sequence, outperforms alternatives across most metrics, particularly demonstrating significant improvement in temporal alignment (DeSync). Additionally, when replacing interleaved-RoPE strategy with conventional RoPE, we observe performance degradation across metrics, confirming that interleaved RoPE effectively enhances audio-video modality alignment. To verify the effectiveness of the unimodal transformer, we further replace unimodal DiT with audio-video dual-stream DiT. The results show that the audio-only transformer achieved superior performance compared with the replacement approach.

\subsubsection{Representation Alignment.}

For representation alignment, we compare two widely-used pre-trained audio self-supervised models: EAT \cite{chen2024eatselfsupervisedpretrainingefficient} and ATST \cite{li2023selfsupervisedaudioteacherstudenttransformer}. Table \ref{tab:6} reveals that using ATST yields the best results, with noticeable improvements in audio quality, temporal alignment, and text-semantic alignment. Notably, combining EAT and ATST leads to performance degradation across most metrics, attributable to the divergence in feature distributions between the two models, which prevents them from providing robust guidance during representation alignment. Furthermore, we investigate the effects of applying REPA in different stages and layers. The results in Table \ref{tab:7} show that REPA achieves optimal performance when applied in unimodal DiT, with additional observations suggesting better outcomes when applied to shallower layers of the unimodal blocks.

\subsection{Discussion}
\subsubsection{Balanced Visual and Textual Semantics.}
The structural innovation of HunyuanVideo-Foley stems from its strategic use of distinct attention mechanisms for visual and textual feature injection. This approach effectively addresses the issue of generated audio relying excessively on text semantics while overlooking video semantics. The experiments show that HunyuanVideo-Foley achieves superior performance across visual-semantic alignment (IB) with maintaining competitive text-semantic alignment (see Section 4.2 and 4.3), which reveals that joint attention is particularly effective for aligning video features with strong temporal correspondence to audio, whereas separate cross-attention better processes text features that convey global contextual information.

\subsubsection{Enhanced Audio Fidelity Through REPA Strategy and Dataset Scaling}
HunyuanVideo-Foley significantly improves the quality of diffusion-based generation by introducing the REPA training strategy. This approach effectively aligns the hidden representations of DiT with robust self-supervised features (see Section 4.3). Additionally, our proposed data pipeline facilitates the scalable construction of high-quality datasets, further enhancing model performance.

\section{Conclusion}
In this work, we present HunyuanVideo-Foley, a novel TV2A framework with REPA strategy that achieves high-fidelity audio generation with balanced alignment of visual dynamics and text context. Meanwhile, we propose an efficient data pipeline, providing robust support for TV2A data scaling. Comprehensive experimental results demonstrate that HunyuanVideo-Foley achieves new SOTA performance in text-video-to-audio generation, particularly excelling in video-semantic alignment, temporal synchronization, and audio quality.

\bibliography{aaai2026}

\clearpage
\appendix
\begin{onecolumn}
\section{Experiment Details}
\subsection{Bandwidth Tagging.}
To address the varying sampling rates in our training data, we introduce a bandwidth tagging strategy. Audio samples with sampling rates above 16 kHz receive a ``high-quality" tag in their captions. During inference, we correspondingly append this tag to all input captions. Our experiments demonstrate that this method successfully conditions the model to associate the ``high-quality" tag with higher sampling rates, resulting in audio outputs with enhanced high-frequency detail preservation. This bandwidth-aware conditioning significantly improves spectral fidelity, as shown by the superior high-frequency retention in the generated waveforms.

\subsection{Evaluation Metrics.}
We adopt a comprehensive suite of metrics spanning multiple dimensions:

\begin{itemize}
    \item \textbf{Distribution Matching}
    \begin{itemize}
        \item \textbf{Fréchet Distance (FD)}: Measures the similarity between generated and real audio feature distributions using mean and covariance statistics (lower values indicate better alignment), computed using PANNs and PaSST embeddings.
        
        \item \textbf{Kullback-Leibler Divergence (KL)}: Quantifies probability distribution divergence between generated and real audio features through PANNs.
    \end{itemize}
    
    \item \textbf{Audio Quality}
    \begin{itemize}
        \item \textbf{Inception Score (IS)}: Evaluates quality and diversity through the PANNs classifier.
        
        \item \textbf{AudioBox-Aesthetics}:
        \begin{itemize}
            \item \textit{Production Quality (PQ)}: Focuses on the technical aspects of quality instead of subjective quality. Aspects including clarity \& fidelity, dynamics, frequencies and spatialization of the audio;
            \item \textit{Production Complexity (PC)}: Focuses on the complexity of an audio scene, measured by number of audio components. In our experiments, we found that audio with significant noise and unnatural artifacts tend to receive higher PC scores, whereas clean, human-perceptually pleasant audio samples are assigned lower scores. Therefore, in the context of Foley generation, we argue that lower PC scores are preferable, as they indicate reduced noise and closer alignment with human auditory perception;
            \item \textit{Content Enjoyment (CE)}: Focuses on the subject quality of an audio piece. It’s a more open-ended axis, some aspects might includes emotional impact, artistic skill, artistic expression, as well as subjective experience, etc;
            \item \textit{Content Usefulness (CU)}: Also a subjective axis, evaluating the likelihood of leveraging the audio as source material for content creation.
        \end{itemize}
    \end{itemize}
    
    \item \textbf{Visual-Semantic Alignment}
    \begin{itemize}
        \item \textbf{ImageBind (IB) Cosine Similarity}: Measures cross-modal alignment between video frames and generated audio embeddings using ImageBind's joint embedding space (higher scores indicate better alignment).
    \end{itemize}
    
    \item \textbf{Temporal Alignment}
    \begin{itemize}
        \item \textbf{DeSync}: Predicts audio-visual synchronization errors via Synchformer (lower values indicate tighter temporal coherence).
    \end{itemize}
    
    \item \textbf{Text-Semantic Consistency}
    \begin{itemize}
        \item \textbf{LAION-CLAP Score}: Measures semantic similarity between input text and generated audio through LAION-CLAP (higher scores reflect better textual grounding).
    \end{itemize}
    
    \item \textbf{Subjective Evaluation}
    \begin{itemize}
        \item \textbf{MOS-Q}: The acoustic quality and auditory naturalness of the generated audio, independent of video content (1-5 scale);
        \item \textbf{MOS-S}: The degree of matching between the category, source characteristics, and physical attributes of the generated audio with the content depicted by the video frames and textual semantics (1-5 scale);
        \item \textbf{MOS-T}: The accuracy of synchronization between the generated audio and visual events, including onset/offset timing and duration (1-5 scale).
    \end{itemize}
    
    \item \textbf{Audio Reconstruction}
    \begin{itemize}
        \item \textbf{PESQ}: Perceptual Evaluation of Speech Quality (1-4.5 scale)
        \item \textbf{STOI}: Short-Time Objective Intelligibility (0-1)
        \item \textbf{SI-SDR}: Scale-Invariant Signal-to-Distortion Ratio (dB)
        \item \textbf{Mel Distance}: Distance between ground-truth and generated Mel-spectrograms
    \end{itemize}
\end{itemize}

\subsection{Baseline Details}
In our experimental setup, we conduct comprehensive comparisons with five baseline models: FoleyCrafter, V-AURA, Frieren, MMAudio, and ThinkSound. To ensure fair comparisons, all models are evaluated using their officially released pre-trained versions, with inference performed on identical hardware configurations and following the original inference scripts. When multiple pre-trained variants are available, we consistently select the best version for benchmarking. Notably, for ThinkSound, we only evaluate the version without Chain-of-Thought (CoT) instructions due to the unavailability for the pre-trained LLM component responsible for generating CoT instructions. A brief introduction to these baselines follows:
\begin{itemize}
    \item \textbf{FoleyCrafter}: A TV2A framework that ensures audio generation through a pretrained text-to-audio model, featuring a semantic adapter with cross-attention for visual relevance, and a temporal controller with onset detection for precise synchronization.
    \item \textbf{V-AURA}: The first autoregressive video-to-audio model achieving fine-grained alignment via high frame-rate visual features and cross-modal fusion.
    \item \textbf{Frieren}: A V2A model based on rectified flow matching for spectrogram generation via ODE sampling. Employs transformer-based cross-modal fusion for alignment.
    \item \textbf{MMAudio}: A TV2A framework jointly trained on video-audio and text-audio data to enhance semantic alignment. Uses flow matching and a frame-level sync module for efficiency, achieving SOTA performance in previous works.
    \item \textbf{ThinkSound}: Integrates Chain-of-Thought reasoning into a three-stage pipeline: foundational Foley generation, interactive object-centric refinement, and language-guided editing.

\end{itemize}

\section{Visualization}
\subsection{Radar Chart.}
To provide an intuitive comparison of evaluation results across different models, we present radar charts in Figure \ref{fig:radar}, which consists of three subplots corresponding to performance evaluations on the Kling-Audio-Eval, VGGSound-Test, and MovieGen-Audio-Bench datasets respectively. The visualization clearly demonstrates that HunyuanVideo-Foley achieves comprehensive performance advantages across all three evaluation benchmarks, consistently outperforming existing methods in almost all metric dimensions. 
\begin{figure*}[h]
    \centering    
    \includegraphics[width=\textwidth]{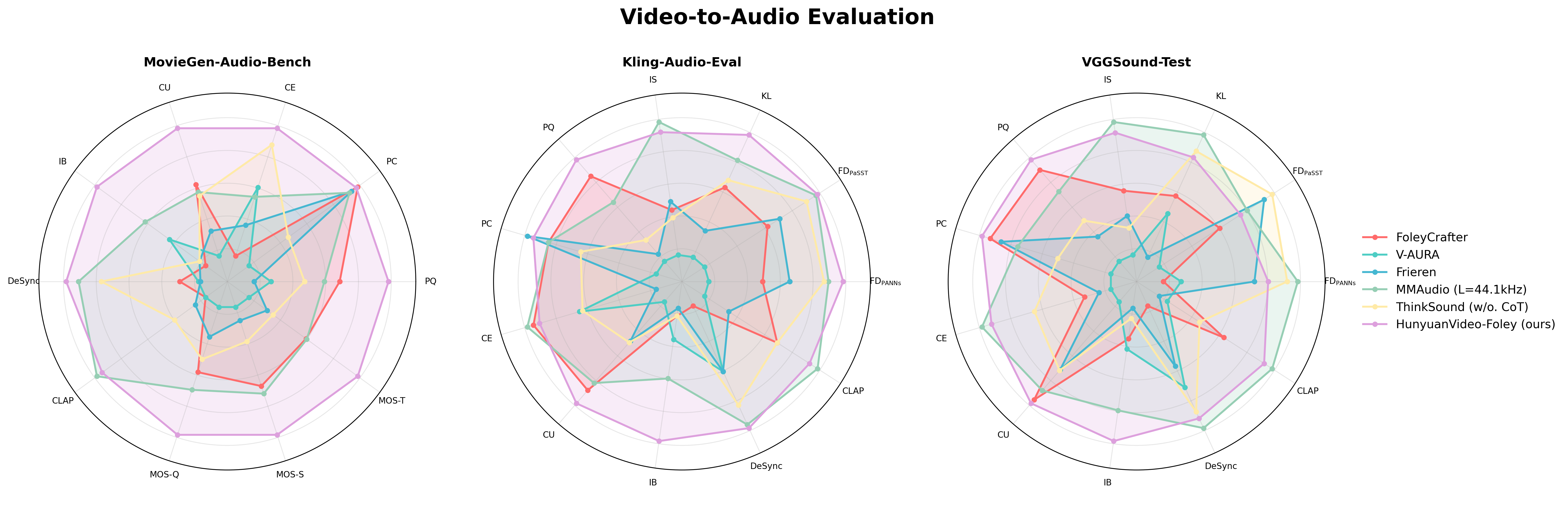}
    \caption{Radar Chart of Video-to-Audio Evaluation. It contains the results on three evaluation set: Kling-Audio-Eval, VGGSound-Test, and MovieGen-Audio-Bench, demonstrating that HunyuanVideo-Foley achieves comprehensive superiority.}
    \label{fig:radar}
\end{figure*}

\subsection{Spectrogram.}
We present spectrogram visualization between our method and existing approaches in Figure \ref{fig:2} and \ref{fig:3}. Notably, our method demonstrates stable performance in preserving high-frequency components without spectral leakage, while maintaining precise temporal alignment between audio events and corresponding actions.
\begin{figure*}[h]
    \centering    
    \includegraphics[width=0.85\textwidth]{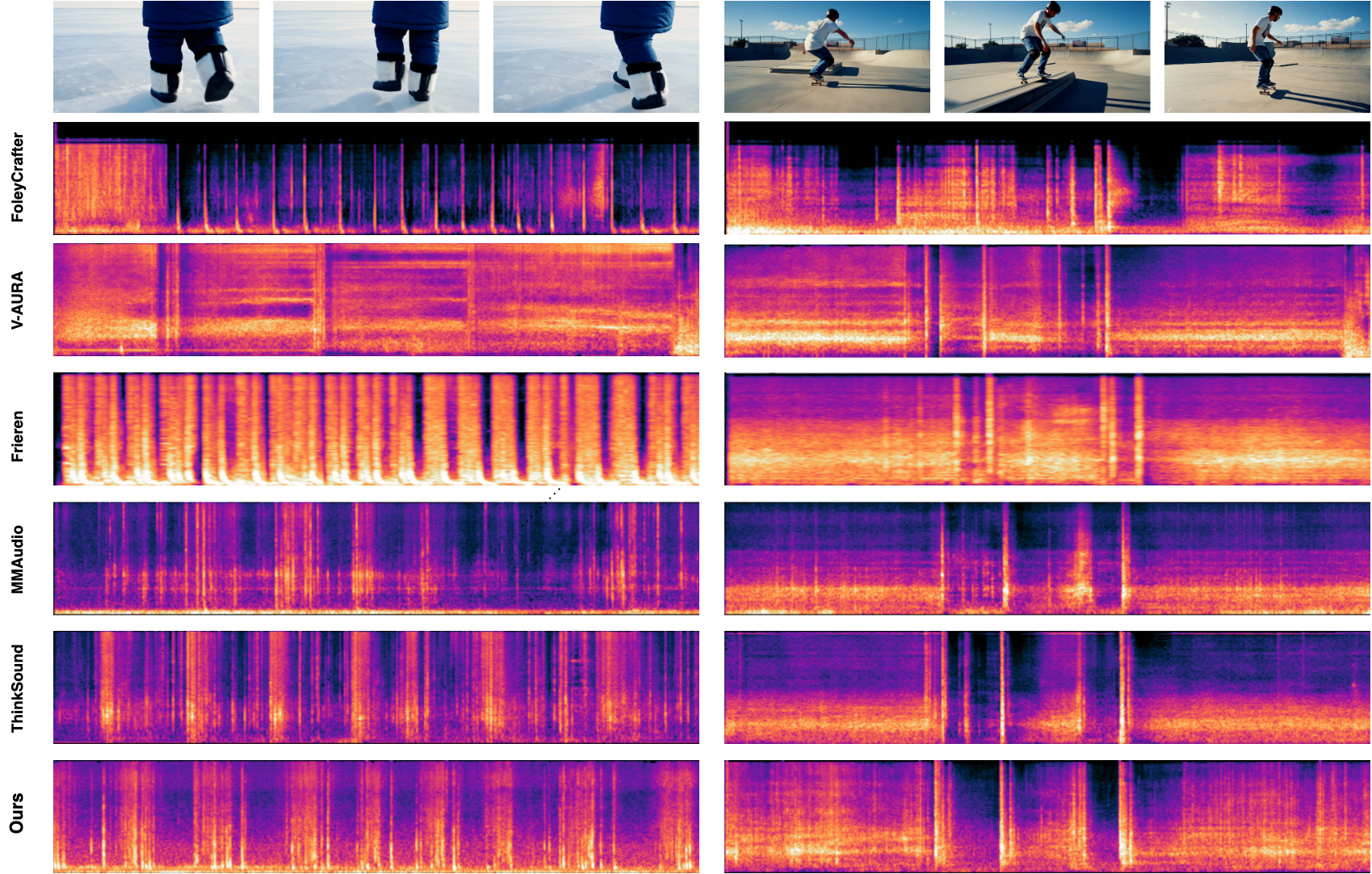}
    \caption{Left: The video sequence illustrates a walking scenario on icy surfaces, where our proposed method achieves precise temporal alignment for both the initiation/termination timing and the duration of each step. Right: Spectral analysis confirms accurate synchronization with the temporal characteristics of human movements in the skateboarding scenario.}
    \label{fig:2}
\end{figure*}

\begin{figure*}[h]
    \centering    
    \includegraphics[width=0.85\textwidth]{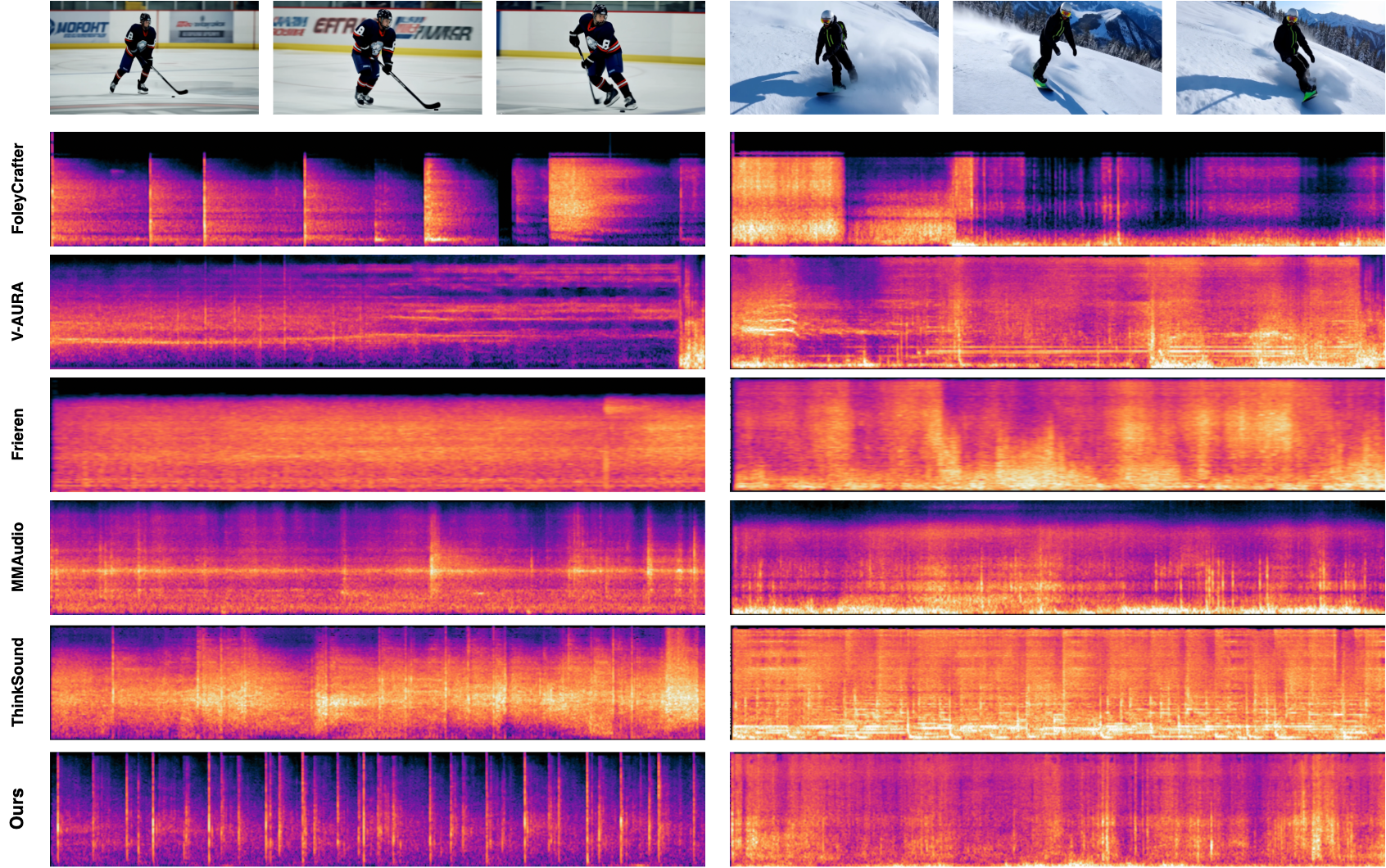}
    \caption{Left: In the ice hockey scenario involving rapid rhythmic auditory cues, our spectral analysis demonstrates robust performance in detecting subtle motion variations synchronized with the sound patterns. Right: Our method preserves the full spectral representation in complex skiing scenario where motion-sound alignment is less distinct, with no discernible degradation of high-frequency components in the spectrogram.}
    \label{fig:3}
\end{figure*}
\end{onecolumn}

\end{document}